\documentstyle[aps,prl,multicol,epsf]{revtex}

\def\ep0{\epsilon^0}

\def\z1{b}

\def\be{\begin{equation}}
\def\ee{\end{equation}}

\begin{document}

\title{Extinction Transition on a Pie }
\author{Nadav M. Shnerb}
\address{Depatment of Physics, Judea and Samaria College, Ariel, Israel 44837}
\date{\today}
\maketitle

\begin{abstract}
Extinction transition of bacteria under forced rotation is analyzed in pie geometry.
Under convection, separation of the radial and the azimuthal degrees of freedom
is not possible, and the linearized evolution operator is diagonalized numerically.
Some characteristics scales are compared with the results of recent experiments,
and the ``integrable'' limit of the theory at narrow growth region is analyzed. 
\end{abstract}
\vspace{0.25in}
\begin{multicols}{2}

The time evolution of bacterial colonies on a petri-dish has been studied recently
both theoretically and experimentally\cite{eshel,wakita,budrene,kudrolly}.
The colony is relatively simple biological system, and its basic component,
an individual bacterium, involves only ``elementary'' biological processes
like diffusion, food consumption, multiplication, death and perhaps some interaction
like chemotaxis. Studies with some bacteria strains have reported a wide variety
of complex pattern formation, in most cases due to competition for food resources
and chemical interaction. With uniform, not-exhaustible background of nutrients
and without the presence of mutations and chemical signaling, these simple strains
are suppose to invade a region of nutrient rich agar in the form of a front
propagating with some typical velocity, known as the  Fisher front \cite{murray,fisher,kpp}.

Biological problems of colony growth in \textit{inhomogeneous} environment and
under forced convection has been modeled recently by Nelson and Shnerb \cite{ns}
and by  Dahmen, Nelson and Shnerb \cite{dns}. These studies have focused on the
spectral properties of the linearized evolution operator, which becomes non-Hermitian
in the presence of convection  \cite{hn}. An experiment  designed to test these
predictions, has been carried out recently by Neicu et al \cite{kudrolly}. 

In the experiment, a colony of \textit{Bacillus subtilis} bacteria is forced
to migrate in order to ``catch up'' with a shielded region on the the 
petri-dish,
where all the other parts of the dish are exposed to an ultra violet (UV) light,
which (under the experimental conditions) makes  the unshilded  bacteria immotile.
It was assumed that the adaptation of the bacterial colony to the moving shielded
region has nothing to do with information processing or mutual signaling in
the colony, but is attributed solely to the combined effect of ``dumb'' diffusion
of individual bacteria and the larger growth rate under the shelter. Theoretically,
it was predicted that the adaptation of the colony to the moving environment
fails as the drift is faster than the Fisher front velocity, and in this case
the colony lags behind the shelter and an extinction transition takes place. 

In order to get the essence of the theory, let us consider a one-dimensional
example, where bacteria are diffusing on a line parametrized by \( x \), and
are subject to some environmental heterogeneity which implies fluctuating growth
rate. If the bacteria diffuses, multiply  and are forced to migrate with
some convection velocity \( v \), the differential equation which describes
the evolution of the colony is,

\begin{equation}
\label{1}
\frac{\partial c}{\partial t}=D\frac{\partial ^{2}c}{\partial x^{2}}+v\cdot \frac{\partial c}{\partial x}+a(x)c-bc^{2}.
\end{equation}
 where \( a(x) \) is the local growth rate. When the hostile environment outside
the ``oasis'' causes the immediate death of any bacteria, and inside the oasis
there is some positive growth rate, 

\begin{equation}
\label{2}
a(x)=\begin{array}{cc}
a & 0\leq x\leq x_{0}\\
-\infty  & elsewhere.
\end{array}
\end{equation}
If there is no drift, the linearized version of this problem is equivalent to
the (imaginary time) evolution of quantum particle in an infinite potential
well, and is determined by the eigenvalues of the evolution operator, which
gives a colony localized on the oasis if it has some minimal width (which scales
like the width of the Fisher front), \( x_{0}>\pi \sqrt{D/a} \). The introduction
of a drift term into (\ref{1}) is compensated by the ``gauge'' of the evolution
(Liouville) operator eigenfunctions 
\begin{equation}
\label{3}
\phi _{n}=\sin (n\pi x/x_{0})\to e^{\pm vx/2D}\sin (n\pi x/x_{0})
\end{equation}
 together with the eigenvalues ``rigid'' shift 
\begin{equation}
\label{4}
\Gamma _{n}(v)=\Gamma _{n}(v=0)-\frac{v^{2}}{4D}.
\end{equation}
 The theory, thus, predicts an extinction transition as all the eigenvalues
of (\ref{4}) becomes negative, i.e., as \( v_{c}=2\sqrt{aD}-O(1/x_{0}) \),
which is the Fisher velocity \cite{Trefethen}. Right above the extinction transition,
only the largest growth eigenvalue (the ``ground state'') is positive, hence
the nonlinear interaction between eigenmodes {[}the term \( -bc^{2} \) in (\ref{1}){]}
are suppressed at the transition, and the analysis is focused on the ground
state of the linearized operator.

In the experiment\cite{kudrolly}, part of a petri-dish was shielded from the
UV source, and then this shield was given a constant \textit{angular} velocity
with respect to the petri-dish. The corresponding convection velocity \( v(r)=\omega r \)
was  chosen to interpolate between zero (at the rotation axis) and about
\( 2v_{c} \) at the edge of the dish. It turns out that the colony indeed fails
to keep rotating with the shield at about half the radius. On the other hand,
the velocity profile for the bacterial density \( c(r,\omega ) \) does not
equilibrate on the time scales of the experiment (\( \sim 3\, days) \). In
this paper, I consider the differences between the one dimensional system (\ref{1},\ref{2})
and the actual experimental setup. In particular, the two-dimensional nature
of the experiment and the effect of \textit{radial} diffusion are considered
explicitly.

In order to capture the essential physics using the simplest geometry, the same
extinction transition is considered on a pie, i.e., a section of the two dimensional
disc {[} See Fig. (1){]}. Although the shielded region of the experiment \cite{kudrolly}
was not in that shape, it turns out that even in this simple geometry there
is a coupling between the radial and the azimuthal degrees of freedom, and the
spectrum becomes ``chaotic'' when convection takes place. The results for
this case are, accordingly, relevant also to the more complicated geometry of
the experiment.

The basic equation for the bacterial growth problem on a non-uniform substrate,
in the absence of mutation and chemical interactions, is \cite{ns}:

\begin{equation}
\label{5}
\frac{\partial c(\bf {x},t)}{\partial t}=D\nabla ^{2}c(\bf {x},t)+a(\bf {x})c(\bf {x},t)+v\cdot \nabla c-bc^{2}.
\end{equation}

 With no convection and homogeneous, positive \( a \) this equation supports
Fisher front propagation with velocity \( 2\sqrt{Da} \). The experimental situation
corresponds to \( D\sim 10^{-6}cm^{2}/s \) and \( a\sim 10^{-3}/s \). The
Fisher velocity is of order \( 0.1-1\mu m/s \), as has been found experimentally.
The Fisher \textit{width}, which is the characteristic scale of spatial correlations
in their system, is \( \sqrt{D/a}\sim 10^{-2}cm \), much smaller than the petri-dish
radius of few centimeters. 

In cylindrical geometry, Eq. (\ref{5}) takes the form,

\begin{equation}
\label{6}
\frac{\partial c(r,\theta ,t)}{\partial t}=D\nabla ^{2}c(r,\theta ,t)+a(\theta )c(r,\theta ,t)+v\cdot \nabla c
-bc^{2},
\end{equation}

{\narrowtext
\begin{figure}[!tbh]
\vspace{-0.15in}
\begin{center}
\leavevmode
\epsfxsize=6.0cm
\epsfbox{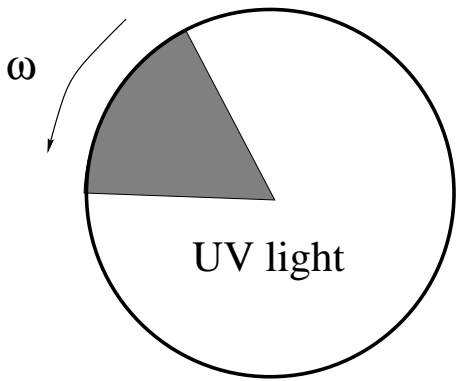} 
\end{center}
\caption{ Pie geometry: the growth rate in the shaded region is \( a, \)
while any bacteria outside this area are  dying instantly due to the UV
light. The shaded region is then rotated at angular velocity  $\omega $. }
\vspace{-0.26cm}
\end{figure}
}
and for rotating petri-dish the convection term is \cite{olami},

\begin{equation}
\label{7}
v\cdot \nabla c=\omega \frac{\partial c}{\partial \theta }.
\end{equation}

Pie geometry is defined by,

\begin{equation}
\label{8}
a(\theta )=\begin{array}{cc}
a & 0\leq \theta \leq \theta _{0}\\
-\infty  & elsewhere
\end{array},
\end{equation}
 i.e., we have  absorbing boundary conditions\cite{pv}:

\begin{equation}
\label{9}
c(r,\theta _{0},t)=c(r,0,t)=0.
\end{equation}
 As for the boundary conditions on the petri-dish edge at \( r=R, \) it is
reasonable to take Von-Neumann boundary and to impose the no-slip condition
on the bacterial density at the surface. However, the data \cite{kudrolly}
seems to indicate extinction of the colony at the edge of the dish. This is
perhaps due to the fact that the width of the boundary layer (which is expected
due to the no slip condition) is about the Fisher width, which has been shown
above to be very small. Accordingly, we further simplify the problem by using, 

\begin{equation}
\label{10}
c(R,\theta \, t)=0.
\end{equation}
 Dropping the term \( -bc^{2} \) at Eq. (\ref{5}), one has the linearized
evolution operator, and for the no-drift \( (\omega =0) \) case, the density
of bacteria at time \( t \) is given by:

\begin{equation}
\label{11}
c(r,\theta \, t)=\sum _{m,n}A_{m,n}e^{a-\Gamma _{m,n}}\phi _{m,n}(r,\theta ),
\end{equation}
 with the eigenstates of the evolution operator,

\begin{equation}
\label{12}
\phi _{m,n}(r,\theta )=\eta _{m,n}J_{\frac{n\pi }{\theta _{0}}}(r/\sqrt{D/\Gamma _{m,n}})\sin (\frac{n\pi \theta }{\theta _{0}}),
\end{equation}
 where \( \eta _{m,n} \) is normalization factor
\begin{equation}
 \eta _{n,m}=\frac{2}{R\sqrt{\theta _{0}}}\frac{1}{J_{[\frac{n\pi }{\theta _{0}}+1]}(R/\sqrt{D/\Gamma _{m,n}})},
\end{equation}
and the constants \( A_{m,n} \) are determined by the initial density distribution
\( c(r,\theta ,t=0). \) The eigenvalues of the Hermitian problem are:

\begin{equation}
\label{13}
\Gamma _{m,n}=D\: \left( \frac{j_{n\pi /\theta _{0}}^{m}}{R}\right) ^{2},
\end{equation}
 where \( j_{n\pi /\theta _{0}}^{m} \) is the m-th zero of the corresponding
Bessel function. 

Let us get an order of magnitude estimate for the time scales which are relevant
to the experiment \cite{kudrolly}. The characteristic times needed for the
``ground state'' to control the system is given by the typical difference
between two eigenvalues. In our case, since the first zeroes of the Bessel functions
are of order 1, the times involved are \( \sim \frac{R^{2}}{D} \). For an experimental
system with \( R\sim 0.01m \) and \( D\sim 10^{-10}m^{2}/s \), the typical
relaxation times are \( O(10^{6}sec)\sim 11\, days \), which is larger than
the typical time of the actual experiment. 

Now I look at the non-Hermitian case, where \( \omega \neq 0. \) Unlike its
Cartesian analogy \cite{ns,dns},  no simple gauge solves the problem,
and separation of variables is impossible. Spanning the space of normalizable
functions by a set of Hermitian eigenstates, the perturbative term \( \omega \partial _{\theta } \)
mixes both quantum numbers \( m \) and \( n \). The matrix elements of the
convection term are:
\begin{equation}
\label{14}
\left\langle n,m|\omega \cdot \partial _{\theta }|k,l\right\rangle =2\omega R^{2}\gamma _{nmkl\; }.
\end{equation}
 where \( \gamma _{nmkl} \) is:
\begin{equation}
\label{15}
\gamma _{nmkl}=\left\{ \begin{array}{cc}
k+n=even & 0\\
k+n=odd & \frac{2kn}{n^{2}-k^{2}}\eta _{n,m}\eta _{k,l} I_{nmkl}
\end{array}\right. 
\end{equation}
and
\begin{equation}
I_{nmkl} = \int ^{1}_{0}J_{\frac{n\pi }
{\theta _{0}}}(j^{m}_{n\pi /\theta _{0}}y)\: J_{\frac{k\pi }{\theta _{0}}}(j^{l}_{k\pi /\theta _{0}}y)\, y\, dy.
\end{equation}
 In order to get the eigenvalues and eigenfunctions at finite angular velocity
one should diagonalize the full non-Hermitian Liouville operator, and the extinction
transition takes place as the ground state (smallest) eigenvalue, \( \Gamma _{1,1} \), becomes
larger than the growth rate \( a \) on the pie. 

As the rotating system is not integrable, it should be studied numerically using
some computer diagonalization of the linearized evolution operator. Essentially,
one should look at the ground state of this operator, since this state dominates
the system close to the extinction transition. 

This numerical analysis, however, may lead to erroneous results if the continuum
limit is not taken carefully. In the most general case, a discretized version
of a model with local growth rate and hopping between sites may be realized
numerically as a matrix, where the growth rates are the coefficients on the
diagonal and the hopping process gives the off-diagonal terms. As any hopping
term is  positive semi-definite, the only negative terms are the local growth
rate, and for any finite matrix, by adding an appropriate multiplication of
the unit matrix, one may get a positive semi-definite matrix with the \textit{same}
eigenvectors. Perron-Frobenius theorem \cite{perron} then implies that ground
state should be a nodeless, positive eigenvector. There is a simple physical
interpretation to this result: since the ground state dominates the system at
long times, and the number of bacteria should not become negative, Perron-Frobenius
theorem should hold. Diagonalizing numerically the evolution operator, one may
get a ground state with nodes, which is physically impossible.

{\narrowtext
\begin{figure}[tbh]
\vspace{-0.4in}
\begin{center}
\leavevmode
\epsfxsize=7.0cm
\epsfbox{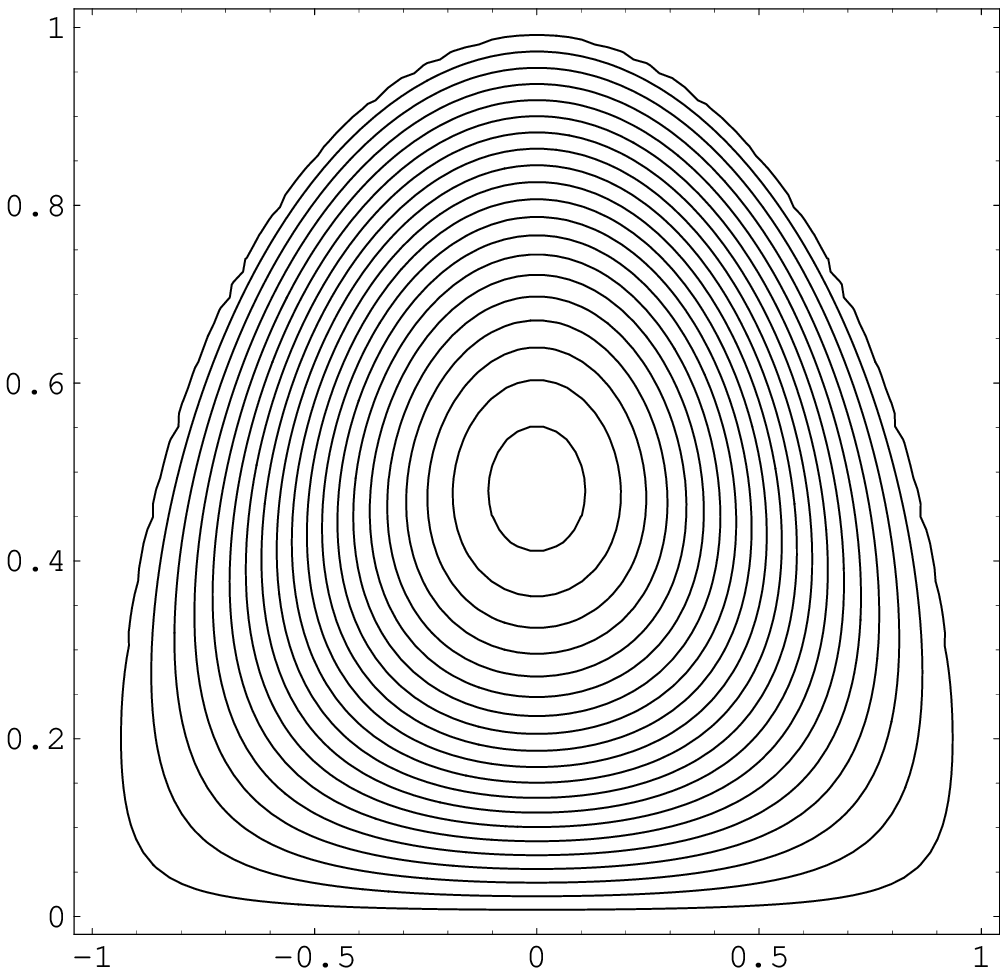} 
\end{center}
\vspace{-0.5in}
\begin{center}
\leavevmode
\epsfxsize=7.0cm
\epsfbox{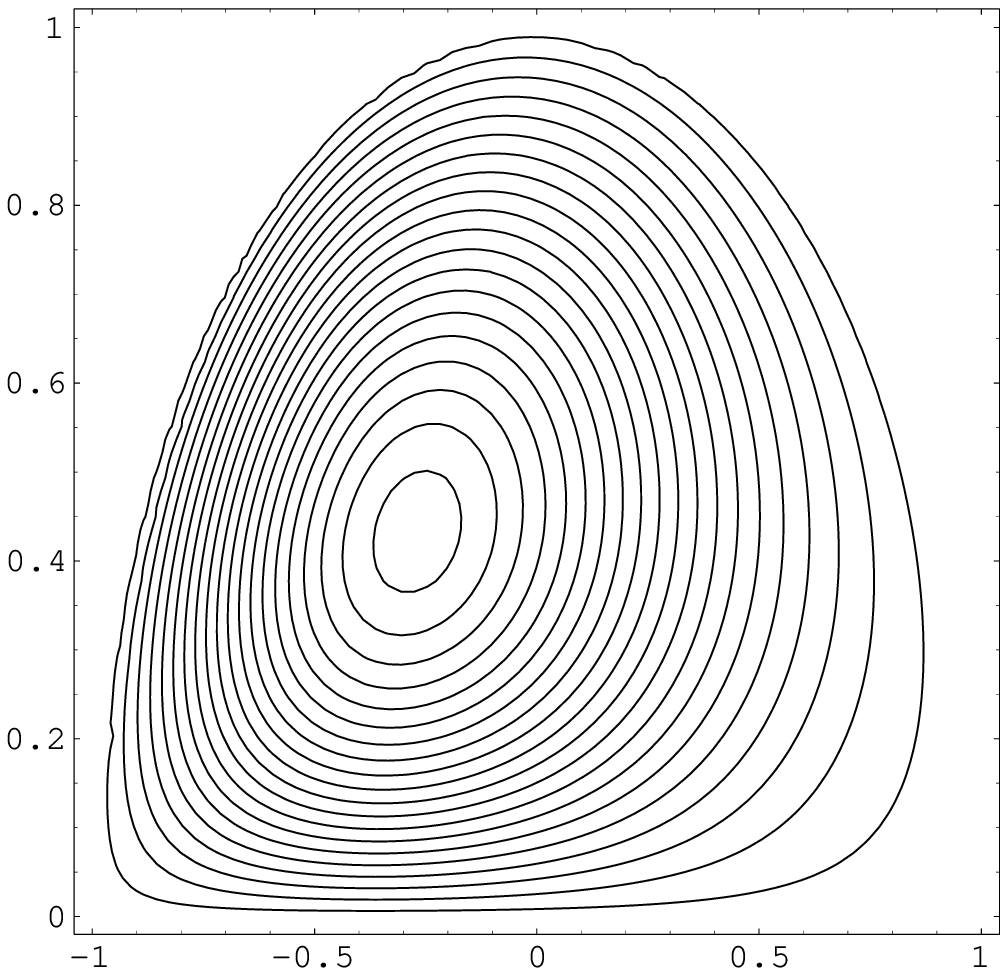} 
\end{center}
\vspace{-0.5in}
\begin{center}
\leavevmode
\epsfxsize=7.0cm
\epsfbox{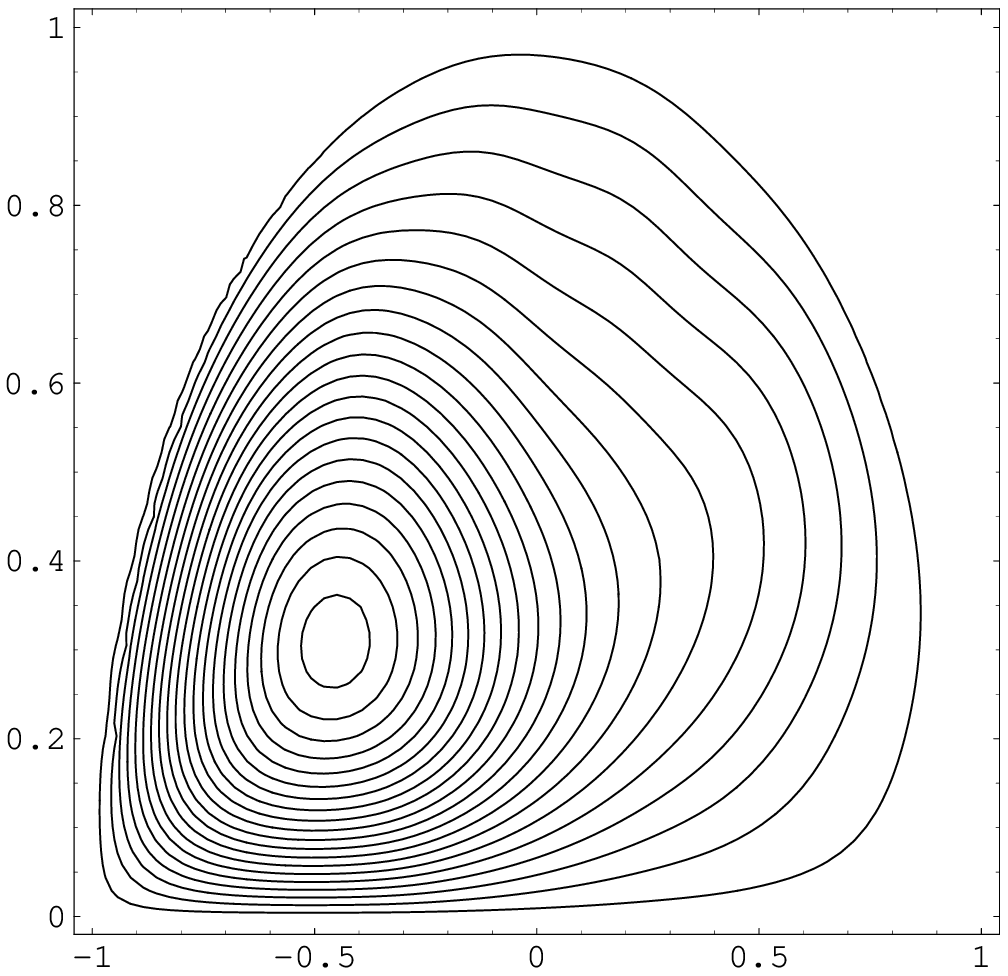} 
\end{center}
 \caption{ Contour Plot of the bacterial density at \( \theta _{0}=\pi  \).
(upper panel) :\( \frac{\omega }{D/R^{2}}=0 \), (middle): \( \frac{\omega }{D/R^{2}}=10 \)
and (lower panel) at \( \frac{\omega }{D/R^{2}}=30. \)}
\vspace{-0.26cm}
\end{figure}
}

{\narrowtext
\begin{figure}[tbh]
\vspace{-0.2in}
\begin{center}
\leavevmode
\epsfxsize=7.7cm
\epsfbox{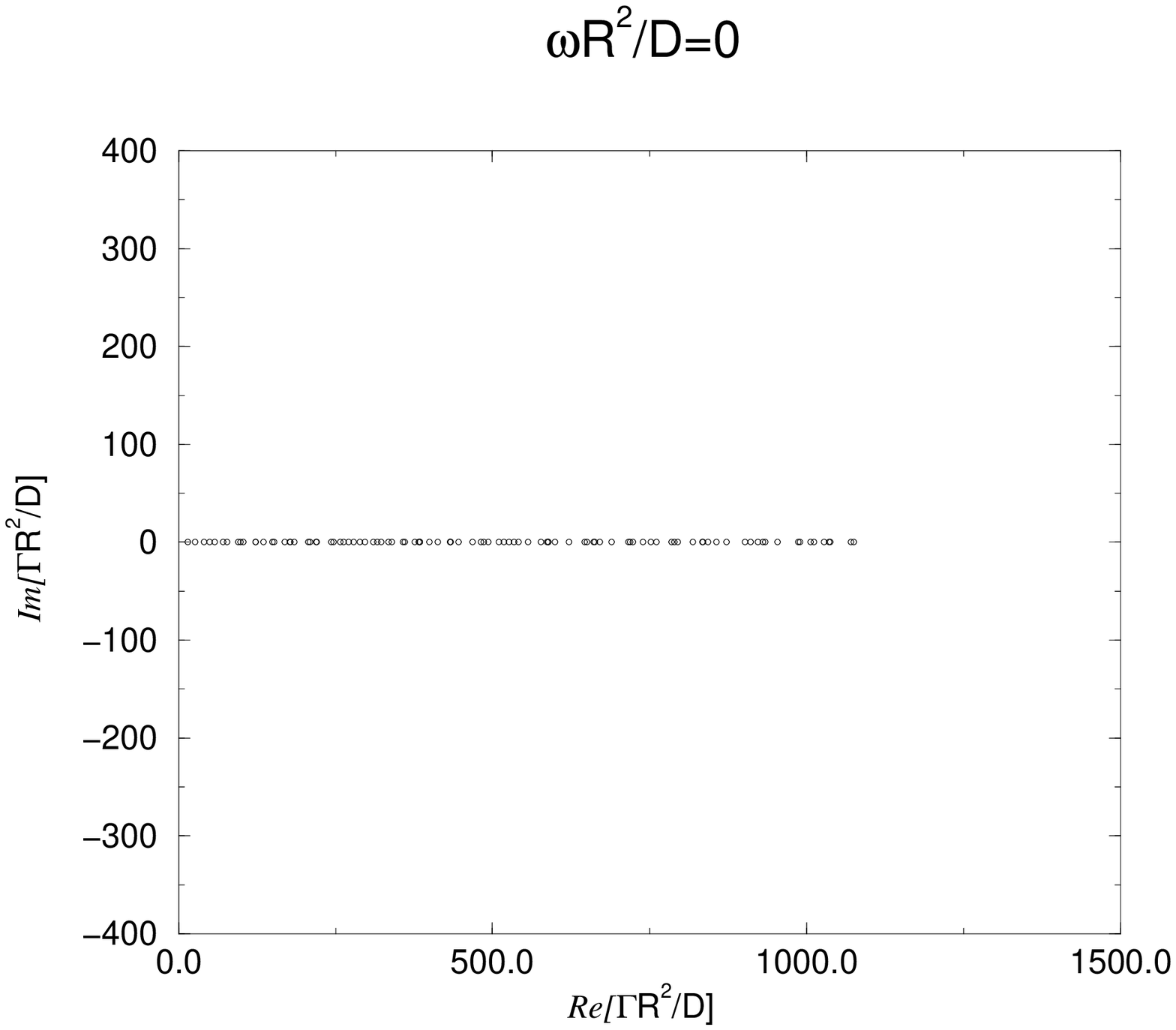} 
\end{center}
\vspace{-0.2in}
\begin{center}
\leavevmode
\epsfxsize=7.7cm
\epsfbox{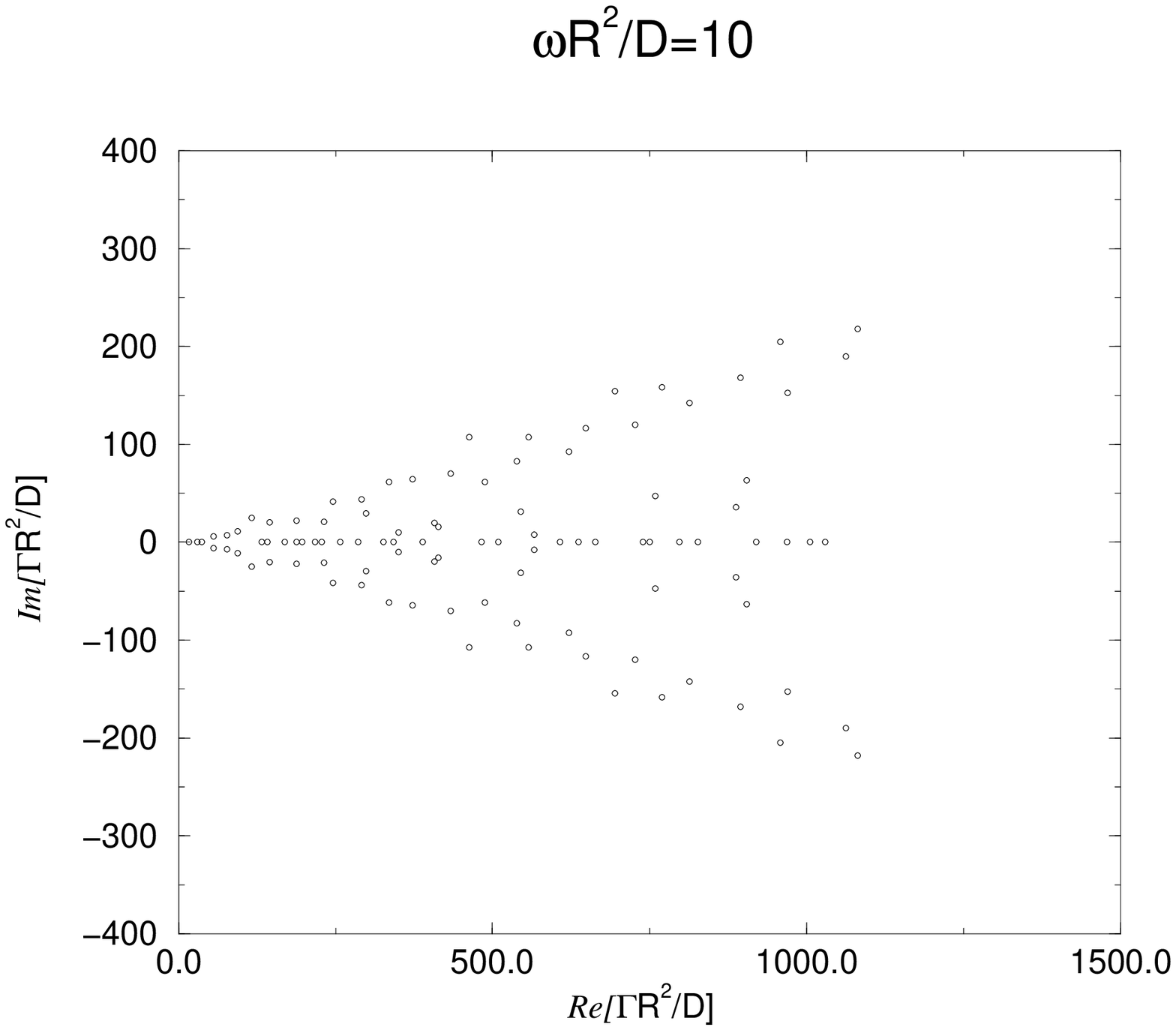} 
\end{center}
\vspace{-0.2in}
\begin{center}
\leavevmode
\epsfxsize=7.7cm
\epsfbox{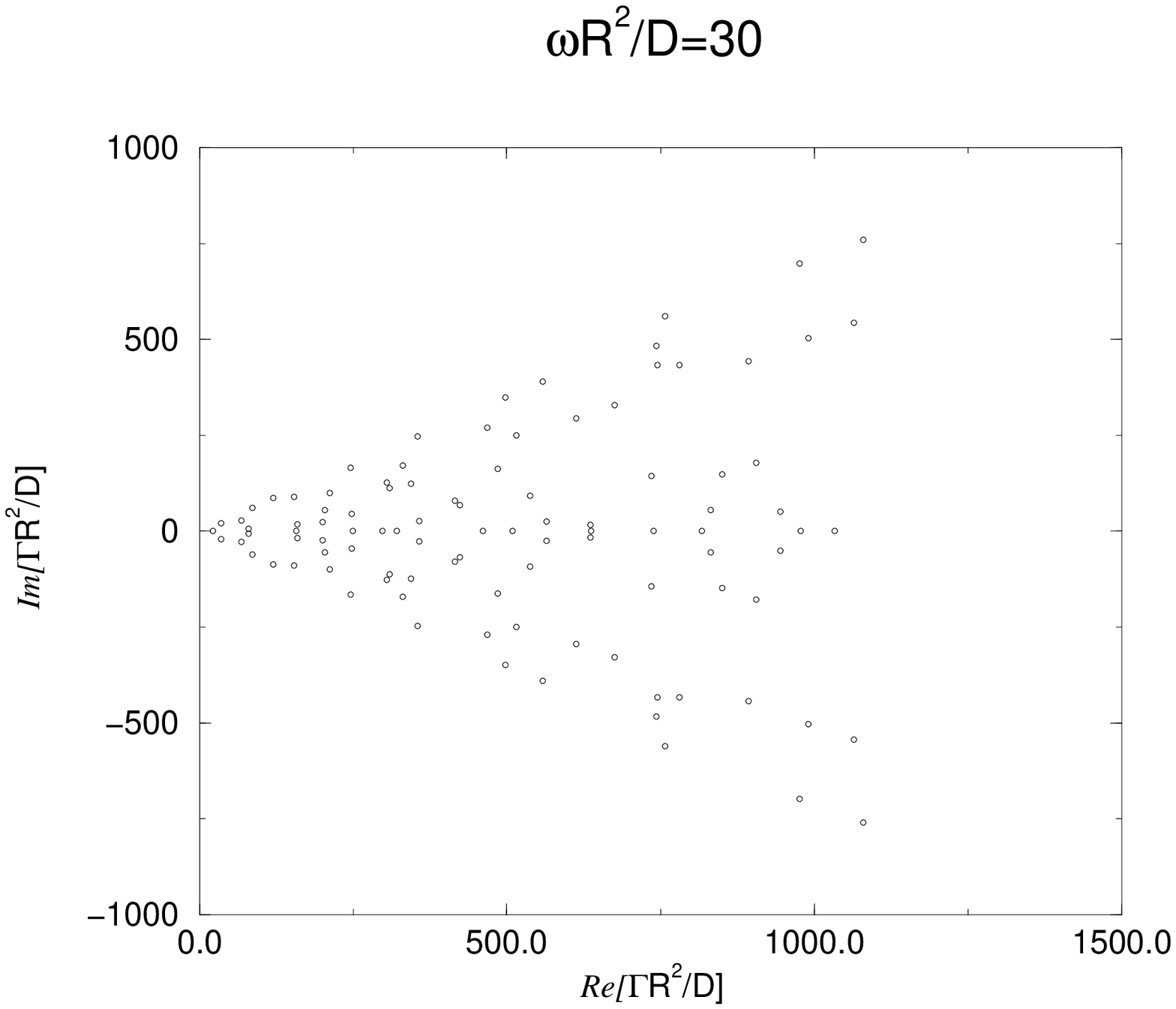} 
\end{center}
 \caption{ First 100 spectral points {[}\( Im(\frac{\Gamma }{D/R^{2}}) \)
vs. \( Re(\frac{\Gamma }{D/R^{2}}) \){]} at \( \theta _{0}=\pi  \). (upper panel) :\( \frac{\omega }{D/R^{2}}=0 \),
(middle): \( \frac{\omega }{D/R^{2}}=10 \) and (lower panel) at \( \frac{\omega }{D/R^{2}}=30. \)}
\vspace{-0.26cm}
\end{figure}
}

In order to solve this problem one should  carefully take  the discrete limit
of the continuum theory. For our case, as \( \theta  \) is discretized to quanta
of \( \Delta \theta  \), the hopping rate due to diffusion becomes \( D/(r^{2}\Delta \theta ^{2}) \)
and the hopping rate due to the drift is \( \pm \omega /\Delta \theta  \).
In order to avoid the (physically impossible) negative hopping rates, one should
keep \( \Delta \theta  \) small enough. If the effective discretization is
given by \( \Delta \theta =\theta _{0}/n \), one should truncate the matrix
(\ref{15}) only as

\begin{equation}
\label{16}
n\sim \frac{\omega R^{2}\theta _{0}}{D}.
\end{equation}

Although (\ref{16}) seems to indicate that the numerical diagonalization of
(\ref{14}) becomes simpler as \( \theta _{0}\to 0, \) this, in fact, is not
the case. As the eigenvalues of the unperturbed problem are related to the zeroes
of the corresponding Bessel functions, and the rotation operator admits matrix
elements only between eigenvalues related to Bessel functions of different order,
it is much simpler to diagonalize (\ref{14}) as \( \theta _{0}>>0 \). As \( \theta _{0}\to 0, \)
the higher \( m \) zeroes of any Bessel of order \( n \) are smaller than
the \( m=1 \) zero of the \( n+1 \) state and the condition (\ref{16}) implies
the diagonalization of an infinite matrix. Accordingly, I present here the numerical
results for the case \( \theta _{0}=\pi . \) This situation does not coincide
with the experimental conditions at \cite{kudrolly}, but there seems to be
no prevention to perform the same experiment with large shielded area.

In Fig. (2),  contour plots of the ground state for different angular velocities
are shown. One may identify clearly the large deviations from the ground state
from its shape at \( \omega =0. \) The largest 100 spectral points for any
case are shown in Fig. (3). 

Fig. (4) presents the ground state eigenvalue, \( \Gamma _{0,} \) in units
of \( D/R^{2}, \) as a function of the angular velocity of the dish. The extinction
transition takes place as this eigenvalue is larger than the growth rate on
the ``pie'', \( \frac{a}{D/R^{2}}, \) as has been found above.

{\narrowtext
\begin{figure}[tbh]
\vspace{-0.15in}
\begin{center}
\leavevmode
\epsfxsize=6.0cm
\epsfbox{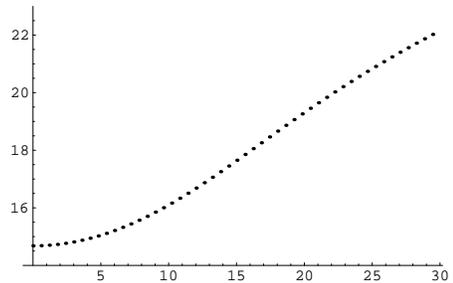} 
\end{center}
\caption {Highest eigenvalue, \( \frac{\Gamma _{0}}{D/R^{2}}, \)
as a function of the angular velocity, \( \frac{\omega }{D/R^{2}}, \) for \( \theta _{0}=\pi . \)
The extinction happens as this eigenvalue is larger than \( a \), the growth
rate inside the pie.}
\vspace{-0.26cm}
\end{figure}
}

{\narrowtext
\begin{figure}[tbh]
\vspace{-0.15in}
\begin{center}
\leavevmode
\epsfxsize=6.0cm
\epsfbox{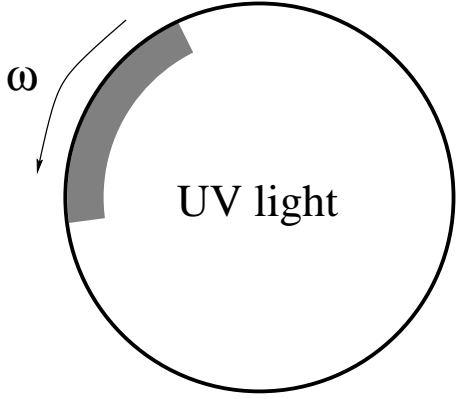} 
\end{center}
\caption{Narrow shell geometry, where the two-dimensional
problem converges to the ``integrable'' case. }
\vspace{-0.26cm}
\end{figure}
}

Let us show now how to get a problem equivalent to (\ref{1},\ref{2}) on a rotating
petri-dish. In order to do that, the geometry should be taken on a narrow shell
as in Fig. (5), i.e., the boundary conditions are,

\begin{equation}
\label{17}
\begin{array}{c}
c(r,\theta _{0},t)=c(r,0,t)=0.\\
c(R_{1},\theta _{0},t)=c(R_{2},0,t)=0.
\end{array}
\end{equation}
 with \( \Delta R=R_{2}-R_{1} \). In the limit \( R_{1}\to \infty  \) at constant
\( n \), the asymptotic expansion of the Bessel functions \( J_{\nu } \) and
\( Y_{\nu } \) at large argument gives the eigenfunctions of the unperturbed
Liouville operator,

\begin{equation}
\label{18}
\phi _{m,n}(r,\theta )\approx \frac{(\Gamma _{n,m}/D)^{1/4}}{\Delta R\sqrt{R_{1\, }\theta _{0}\, }}\sin (\frac{m\pi r}{\Delta R}+\alpha _{n})\sin (\frac{n\pi \theta }{\theta _{0}}),
\end{equation}
 where the phase \( \alpha _{n} \) ensures the boundary conditions at \( R_{1} \)
and the eigenvalues, \( \Gamma _{n,m}=D\frac{m^{2}\pi ^{2}}{(\Delta R)^{2}} \),
are \textit{independent} of \( n. \) The matrix elements of the operator \( \omega \partial _{\theta } \)
are given by

\begin{equation}
\label{19}
\left\langle n,m|\omega \cdot \partial _{\theta }|k,l\right\rangle =\omega \delta _{m,l\, }\gamma _{nk\; }.
\end{equation}
 with

\begin{equation}
\label{20}
\gamma _{nk}=\left\{ \begin{array}{cc}
k+n=even & 0\\
k+n=odd & \frac{2kn}{n^{2}-k^{2}}
\end{array}.\right. 
\end{equation}
where the approximation 

\begin{eqnarray}
\label{21}
\int ^{R_{2}}_{R_{1}}\sin (\frac{m\pi r}{\Delta R}+\alpha _{n})\sin (\frac{l\pi r}{\Delta R}+\alpha _{k})\, \sqrt{r}\, dr\: \nonumber \\ \sim \: \sqrt{R_{1}}\int ^{R_{2}}_{R_{1}}\sin (\frac{m\pi r}{\Delta R}+\alpha _{n})\sin (\frac{l\pi r}{\Delta R}+\alpha _{k})\, dr
\end{eqnarray}
 for \( \frac{\Delta R}{R_{1}}<<1 \) has been used. Accordingly, for any \( m \)
sector, both the diagonal and the off diagonal matrix elements are identical
with the corresponding one dimensional problem, and the results should be the
same. 

In conclusion, the mathematical problem which corresponds to the experiment
by \cite{kudrolly} has been found to be non-integrable, and no simple gauge
transformation connects the eigenvectors of the static and the dynamic problems.
The actual critical velocity and ground state properties have to be studied numerically,
and the limit of very narrow sector (\( \theta _{0}\to 0) \) involves diverging
numerical load. The time scales needed for the ground state to dominate the
system are larger than the duration of the actual experiment, and this explains
the observed inequilebration. 

I wish to thank A. Kudrolli, D. R. Nelson and K. Dahmen for helpful discussions
and comments.

\end{multicols}
\end{document}